\documentclass[conference]{IEEEtran}
\usepackage{cite}
\usepackage{amsmath,amssymb,amsfonts}
\usepackage{algorithmic}
\usepackage{graphicx}
\usepackage{balance}
\usepackage{textcomp}
\usepackage{underscore}
\newcommand{\pp}[1]{\textcolor{black}{#1}}
\usepackage{color}

\usepackage[export]{adjustbox}
\usepackage{float}
\usepackage{enumitem}
\usepackage{verbatim}
\usepackage{booktabs}
\usepackage{array}

\begin{document}
\bstctlcite{IEEEexample:BSTcontrol}
\title{Wireless Environment as a Service\\ Enabled by Reconfigurable Intelligent Surfaces:\\ The RISE-6G Perspective\\
\thanks{This work was supported by the European Union H2020 RISE-6G Project.}}

\author{
    Emilio Calvanese Strinati\IEEEauthorrefmark{1}, 
    George C. Alexandropoulos\IEEEauthorrefmark{2},   
    Vincenzo Sciancalepore\IEEEauthorrefmark{3},  
    Marco Di Renzo\IEEEauthorrefmark{4},\\
    Henk Wymeersch\IEEEauthorrefmark{5},
    Dinh-Thuy Phan-huy\IEEEauthorrefmark{6},
    Maurizio Crozzoli\IEEEauthorrefmark{7},
    Raffaele D'Errico \IEEEauthorrefmark{1},
    Elisabeth De Carvalho\IEEEauthorrefmark{8},\\
    Petar Popovski\IEEEauthorrefmark{8},
    Paolo Di Lorenzo\IEEEauthorrefmark{9},
    Luca Bastianelli\IEEEauthorrefmark{9},
    Mathieu Belouar\IEEEauthorrefmark{10},
    Julien Etienne Mascolo\IEEEauthorrefmark{11},\\
    Gabriele Gradoni\IEEEauthorrefmark{12},
    Sendy Phang\IEEEauthorrefmark{12},
    Geoffroy Lerosey\IEEEauthorrefmark{13},
    Beno\^{i}t Denis\IEEEauthorrefmark{1} \\
     \IEEEauthorrefmark{1}CEA-Leti, France, \IEEEauthorrefmark{2}National and Kapodistrian University of Athens, Greece, \IEEEauthorrefmark{3}NEC Laboratories Europe,\\ \IEEEauthorrefmark{4} Universit\'e Paris-Saclay, CNRS, CentraleSup\'elec, Laboratoire des Signaux et Syst\`emes, France,\\ \IEEEauthorrefmark{5}Chalmers University of Technology, Sweden, \IEEEauthorrefmark{6}Orange Labs, France, \IEEEauthorrefmark{7}TIM, Italy\\ \IEEEauthorrefmark{8}Aalborg University, Denmark, \IEEEauthorrefmark{9} CNIT, Italy, \IEEEauthorrefmark{10}SNCF, France \IEEEauthorrefmark{11}CFR, Italy\\
     \IEEEauthorrefmark{12}University of Nottingham, UK, \IEEEauthorrefmark{13}Greenerwave, France 
}

\maketitle
\begin{abstract}
The design of 
6th Generation (6G) wireless networks points towards 
flexible connect-and-compute technologies capable to support innovative services and use cases. Targeting the 2030 horizon, 6G networks are poised to pave the way for sustainable human-centered smart societies and vertical industries, such that wireless networks will be transformed into a distributed smart connectivity infrastructure, where new terminal types are embedded in the daily environment. 
In this context, the RISE-6G project aims at investigating innovative solutions that capitalize on the latest advances in the emerging technology of Reconfigurable Intelligent Surfaces (RISs), which offers dynamic and goal-oriented radio wave propagation control, enabling the concept of the \textit{wireless environment as a service}. The project will focus on: \textit{i}) the realistic modeling of RIS-assisted signal propagation, \textit{ii}) the investigation of the fundamental limits of RIS-empowered wireless communications and sensing, and \textit{iii}) the design of efficient algorithms for orchestrating networking RISs, in order   
to implement intelligent, sustainable, and dynamically programmable wireless environments enabling diverse services that go well beyond the 5G capabilities. RISE-6G will offer two unprecedented 
proof-of-concepts for realizing controlled wireless environments in near-future use cases. 
    
\end{abstract}
\section{Introduction}
The 5th Generation (5G) of wireless networks is at an early deployment stage, providing a single platform for a variety of services and vertical applications \cite{Parkvall2020}.
However, novel concepts as well as new services and related use cases, incorporating new enabling technologies to satisfy future needs \cite{Calvanese2020}, are already being identified for addressing the current predictions for the performance requirements of the next 6th Generation (6G) of connect-and-compute networks around $2030$. 
Among those predictions belong the up-to-10 Gbps/m$^3$ capacity, 100$\mu$s latency, 1 Tb/J energy efficiency, and 1 cm localization accuracy in 3 Dimensions (3D) \cite{Calvanese2019}, which will need to be offered individually or in various combinations.

As a common view, wireless networks are designed considering the signal propagation environment as a black box that cannot be artificially controlled. 
This would exacerbate the simultaneous fulfilment of fundamentally conflicting targets, such as boosting quality of experience for different types of services at different places of \textit{intended users} (e.g., ultra-high data rate in hotspots, extreme reliability in factories, and massive connectivity in ultra dense areas), limiting the signal to \textit{non-intended users} (e.g., users' ElectroMagnetic Field Exposure (EMFE) and eavesdroppers), and limiting the network and device energy consumption. Recently, there has been a surge of interest in Reconfigurable Intelligent Surfaces (RISs)~\cite{huang2019reconfigurable} as hardware-efficient and highly scalable means to realize desired dynamic transformations on the signal propagation environment in wireless communications~\cite{liaskos2018new,di2019smart}. RISs are man-made surfaces with thousands of reconfigurable elements, which can support various functionalities (e.g., control multipath geometry for localization, limit EMF exposure, mitigate obstructions, and extend radio coverage in dead zones). The RIS technology is envisioned to coat objects in the wireless environment~\cite{huang2019holographic} (e.g., building facades and room walls), and can operate either as a reconfigurable beyond-optics reflector \cite{WavePropTCCN}, or as a transceiver when equipped with active transmit~\cite{DMA2020} and receive~\cite{hardware2020icassp} radio-frequency elements. 

The modeling of signal propagation in such novel RIS-empowere conditions appears very challenging~\cite{DiRenzoApr2020, risTUTORIAL2020} involving a number of algorithmic solutions.
In~\cite{Gradoni2020}, 
a free-space model based on an impedance matrix formalism for RISs with discrete elements has been presented, while the corresponding
%
optimization of the RIS settings for communication purposes is largely disussed in~\cite{huang2019reconfigurable,risTUTORIAL2020}. 
Additional research directions focus on the challenging task of channel estimation in RIS-empowered communication systems, where
multiple users and multiple RISs with large numbers of elements and non-linear hardware characteristics are considered~\cite{JSAC2020risma}. The relevant literature includes pilot-assisted cascade channel estimation approaches \cite{ZhangOFDM} as well as deep learning frameworks \cite{HuangDNN} that overcome the need for explicit channel estimation. 

RISs will also have applications beyond communication, as in localization and sensing. 5G systems can make use of time, power, and angle measurements to accurately estimate the position and orientation of user devices \cite{del2017survey}. Localization algorithms can also turn received secondary paths in multipath channels into an advantage by using their estimated location-dependent radio parameters as a source of information. 
This 
improves localization continuity in highly obstructed scenarios, whenever a map/layout of the environment or a-priori parametric semi-deterministic models are available. If not, the map should be estimated using   Simultaneous Localization And Mapping (SLAM) methods. However, even in such multipath-aided cases, the exploited ElectroMagnetic (EM) interactions of the propagation environment remain uncontrolled. Despite its inherent advantages, the use of RIS for localization has received very limited treatment up to date. An overview of main research challenges of RISs from the perspectives of both localization and mapping was provided in~\cite{wymeersch2020radio}.

Beyond the above-mentioned state-of-the-art on the RIS technology, this paper highlights the RISE-6G project's perspective that will build a novel wireless connectivity paradigm
empowered by RISs with a two-fold advantage:
$i$) enable highly concentrated in time and space service delivery to intended end users and $ii$) remove energy from indoor/outdoor regions where non-intended users are present. 
This paradigm will pioneer the concept of the \emph{wireless environment as a service}, which will promote dynamical trade-off of high capacity connectivity, Energy Efficiency (EE), EMFE, localization and sensing accuracy, and secrecy guarantees over eavesdroppers, while accommodating specific regulation on spectrum use, data protection, and EMF emission. 


\section{RIS-Empowered System Architecture}
\label{sec:RISSystem}


Traditionally, a wireless network infrastructure is based on elements that operate at the PHY/MAC/link layers and above (e.g., base stations, access points) or even Layer 1 only (e.g., repeaters). Differently from this, an RIS which controls the propagation environment can be considered to operate at Layer 0. As such, RIS brings new directions in the evolution of the wireless architecture, whose investigation is one of the overarching objectives of the RISE-6G project. In this section, use cases that can benefit from deploying RISs are introduced with their associated Key Performance Indicators (KPIs) and the related network architectures and deployment strategies.

\subsection{Scenarios and Use Cases}\label{subsec:T2.1}

\begin{figure}
    \centering
    \includegraphics[width=\columnwidth]{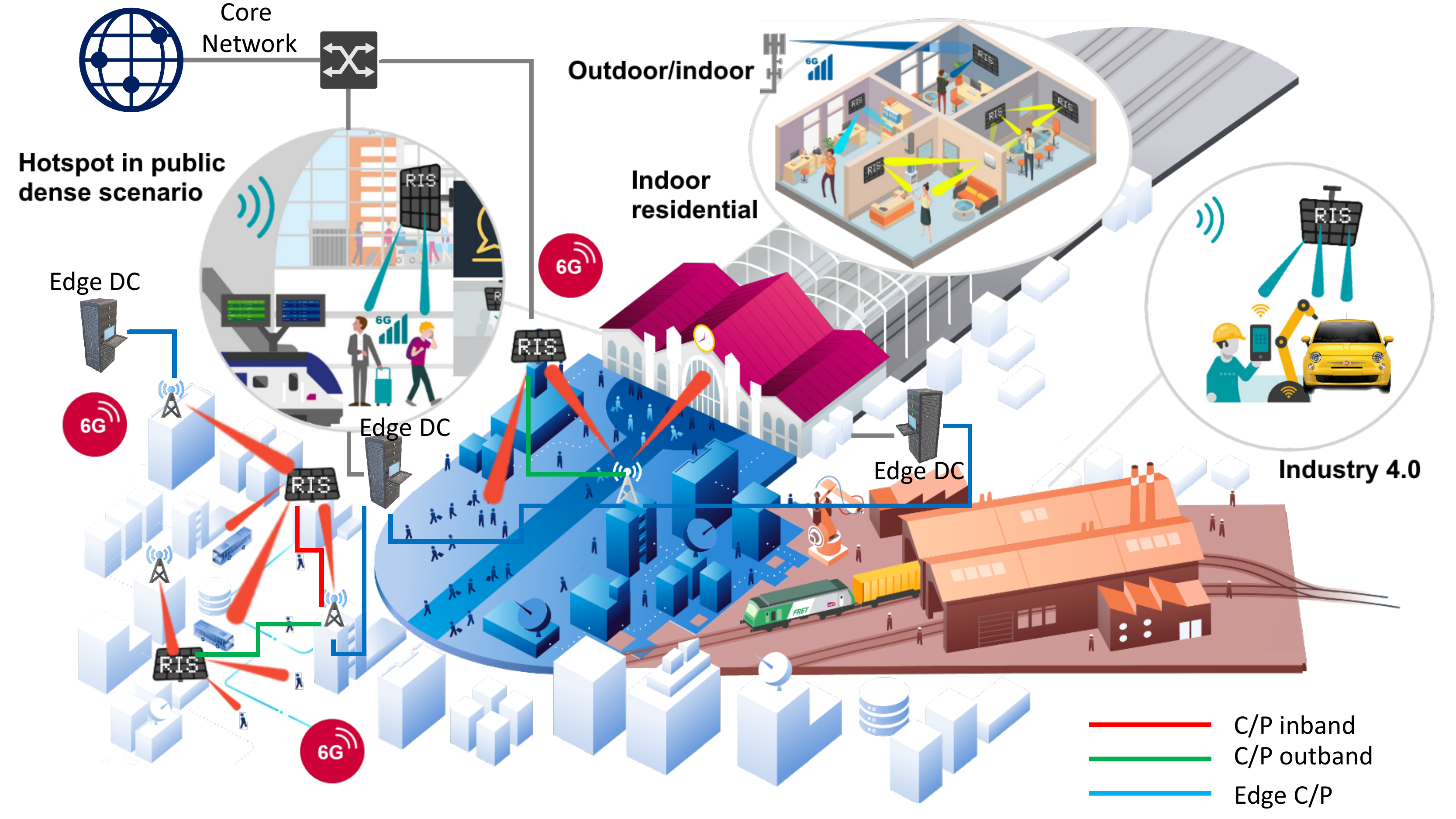}
    \caption{The RISE-6G system architecture showing the inband and outband Control Planes (C/P) among RISs and the edge Data Centers (DC). 
    }
    \label{fig:arch}
\end{figure}

\pp{Considering both sub-6GHz and millimeter-Wave (mmWave) frequency bands, several scenarios and use cases will be identified and analyzed
according to the following steps: \emph{i)} specification of a reference scenario; \emph{ii)} technological and economical analysis of use cases that can benefit from RIS operation; \emph{iii)} analysis of the impact on spectrum and EMFE; \emph{iv)} detailed specification of a use case that favorably illustrates the use of RISs in a given scenario.}

\pp{Among the scenarios that are likely to benefit from the use of RISs, the following will be prioritized:}
\begin{itemize}
    \item \pp{\emph{Enhanced connectivity and reliability.} If the Quality of Experience (QoE)/Quality of Service (QoS) of a wireless network is below expectation, an RIS is applied to improve the radio coverage both in outdoor and indoor. An RIS needs to be implemented in compliance with regulations and EMFE limits;}
    \item \pp{\emph{Localization and sensing.}} The RIS technology is expected to enable advanced sensing and localisation techniques for environment mapping, motion detection, opportunistic channel sounding and passive radar capabilities applied to various environments:  industrial, high user-density and indoor; and
    \item \pp{\emph{Sustainability and security}.} RIS-empowered networks are expected to enable the reduction of the energy spent to radiate the wireless signals, so as to improve the EE, reduce the EMFE, and increase the security due to highly directive and location-dependent communications.
\end{itemize}

\pp{Two exemplary scenarios that have been selected for RIS-empowered networks are a train station and a vertical Industry 4.0 setup, respectively.}

\subsection{Metrics and KPIs for RIS-Empowered Wireless Systems}\label{subsec:T2.2}
Relevant performance metrics and target KPIs to be used by the technical work packages will be identified and defined, and then re-assessed depending on the findings, in order to assess the performance of RIS-empowered wireless systems for the three types of scenarios defined in~Sec.\ref{subsec:T2.1}: from the traditional coverage/capacity metrics, such as data rate and delay, to a set of KPIs related to localization and sensing, such as spatial accuracy, localization availability, and mapping latency. Furthermore, existing metrics on EE (delivered data versus spent energy) will be supplemented by two other ones: \textit{i}) Spectral Secrecy Efficiency (SSE), initially defined as the spectral efficiency attained at the intended user normalized by the cost in terms of data rate successfully captured by the eavesdropper; and \textit{ii}) EMFE Efficiency (EMFEE), defined as the data rate attained by intended users normalized by the cost in terms of exposure of the intended and non-intended users.

In the project's vision, an RIS-empowered network will be able to dynamically generate ``QoE boosted areas" where customers can get guaranteed QoE, ``localization boosted areas" where they can enjoy high precision localization, ``SSE boosted areas" where they can enjoy secured communications, ``EMFEE boosted areas" where environment-conscious customers can avoid EMFE, ``EE boosted areas" where they can improve their energy EE footprint. In such a way, the ``boosted performance" concept, which do not exist today in this context, will be defined and implemented in areas identified by service providers on the basis of marketing insights. Potential trade-offs will be proposed when multiple targets and conflicting KPIs are found. A high-speed train station will be one of the primary use cases where this concept will be applied.

\subsection{Network Architectures and Deployment Strategies}\label{subsec:T2.3}
Suitable network architectures will be specified by leveraging on the technology components developed within the project, as described in the subsequent sections. This will be done for each scenario defined in~Sec.\ref{subsec:T2.1}, according to metrics and KPIs defined in~Sec.\ref{subsec:T2.2}.
Depending on scenarios and application needs, flexible RIS devices will be organized in a network of space-time reconfigurable mirrors, orchestrated together by advanced algorithms running on ad-hoc controllers so as to dynamically change the propagation settings of the installed RISs, based on real-time/predicted network dynamics.

A key new element of the network architecture is the communication channel through which an RIS is controlled and reconfigured. There are two generic ways for RIS control: \emph{i)} in-band control, where the properties of the RIS are dynamically configured by using the same wireless signals whose propagation is affected by the RIS; \emph{ii)} out-of-band control (green connectors on Fig.~\ref{fig:arch}), which takes place over a wired or wireless communication interface not affected by the RIS operation. For each scenario we will determine the appropriate type/combination of RIS control channel to be used.


The most suitable RIS-empowered network architectures and deployment strategies will be selected to be at least partly implemented in a trial, as will be described in Section~\ref{sec:trial}.

\section{Research on Technology Components}
\subsection{Modeling, Design, and Characterization}\label{sec:modeling}
Advanced EM-compliant and environment-aware RIS models will be developed within the RISE-6G project in order to predict the operation of the RIS in the scenarios of interest.
The continuous models provided by metasurface theory will be extended to capture the effect of the tunable discrete structure of the RIS. Inherently, the fundamental research questions concerning RIS modeling are centered around the ability to include: \textit{i}) the mutual coupling between unit cells for arbitrary cell topology and tuning circuity within; \textit{ii}) the near-field interaction between transmitter/receiver and RIS; and \textit{iii}) the integration of RISs within the overall multi-path propagation channel.

RISE-6G will consider impedance models that capture the scattered field behaviour of the discrete surface structure as well as the interaction among different close surfaces in multi-RIS scenarios. Transmit and receive antenna arrays will be also considered in the impedance formalism so that an end-to-end channel model can be constructed between port voltages and currents. These impedance models will rely, on one hand, on the antenna theory with canonical elements, on the other hand, they will exploit full-wave simulation of the unit cells (hybrid model). 
These models will be exploited for design and evaluation of RISs in ideal free-space conditions. 


Rich multi-path fading can be captured by a purely statistical model, the Random Coupling Model (RCM) \cite{Ma2020}, obtained for an antenna impedance operating in highly reflective, large, and complex environments. Deterministic tools will be used to predict the RIS-aided channel propagation. 
An energy flow method that computes densities of rays on numerical meshes, the Dynamical Energy Analysis (DEA), will be extended to include the RIS effect on multi-path propagation. 
The method represents ray densities in a joint space of position and direction of travel, the so called phase-space, through the Wigner function technique \cite{Gradoni20181}. 
An averaging procedure will embed the phase-space density of the RIS scattered field into DEA. 
A model of the RIS scattered field obtained from the Green function approach will be used to integrate their behaviour within the DEA approach in order to have a powerful and unique tool for including the RIS within coverage planning tools.
The model validation will be done through a rich activity of EM characterization of both RIS and the propagation environment itself \cite{Mudonhi21}.

Different RISs (transmittive and reflective) will be designed in RISE-6G, spanning from sub-6 GHz to D-band. Different technologies (PIN-diodes, varactors, and MEMS) will be considered for unit-cell tunabilty, as well as realize space and polarization configurability. Depending on the selected frequency bands, different architectures will be proposed with full or discrete phase control. Large RIS structures with more than $400$ unit elements will be fabricated. 


\subsection{RIS for Enhanced Connectivity and Reliability}

A look at the current state-of-the-art in the RIS research paints a research landscape in which there is a wide range of assumptions. Those assumptions include different models for the surface, the propagation channel, and communication settings. This variety of assumptions makes it difficult to draw meaningful conclusions in terms of performance limits or system design choices. As advocated in RISE-6G, this observation calls for a rigorous framework, encompassing: a) an accurate communication model; b) an account of the control signaling necessary for RIS operations; and c) a performance evaluation framework assessing the essential features of RIS-empowered communications. The majority of the current research adopts simple communication models that fit far-field communications and incorporate  simplified models of the surface \cite{risTUTORIAL2020}. However, the way a metasurface affects wireless propagation properties depends on many characteristics, including the properties of the surface and its unit elements, as well as their interactions \cite{Gradoni2020}, their distance to the transmitter and receiver, and the characteristics of the channel. RISE-6G will develop communication models that are accurate and mathematically tractable, and that account for the circuit design of the surface and its EM properties. Such models are essential to design and optimize RIS-empowered systems, but also to assess their performance limits, their advantages, and  limitations. Based on those accurate communication models, RISE-6G will investigate the connectivity and reliability properties of RIS-empowered systems, providing fundamentals in communication and assessing achievable performance.
The project will also focus on the design and optimization of control signaling protocols, channel estimation methods, and RIS settings, as well as resource allocation and scheduling techniques to enable an effective and reliable support of RISs for seamless distributed smart end-to-end support of communication. Furthermore, RISE-6G will determine the best strategies according to communication metrics for an easy and flexible deployment of RISs considering boosted connectivity areas. 

Mobile Edge Computing (MEC) will also be deeply investigated within the RISE-6G project. In particular, RISE-6G will design a dense network of RISs to enhance the performance of MEC systems, with the aim of delivering pervasive, low-latency, and energy-efficient edge services to mobile users. The key research question is: \emph{How to exploit and design RISs in MEC to strike the best possible trade-off between energy, latency, and accuracy of edge computing tasks?} The project will achieve this goal by jointly optimizing radio, computation, and RIS resources to enable energy-efficient edge computing, while keeping the overall QoS (i.e., delay, accuracy) within a prescribed value dictated by the specific application, and adapting on the fly to dynamic scenarios (e.g., mobility, blocking events). Machine learning will play a key role in RISE-6G designs, enabling to learn how to optimize edge resources over time, decide which users should be served through the available RISs, and provide automation and self-healing capabilities for optimal operational decision.
\begin{figure}
    \centering
    \includegraphics[width=0.9\columnwidth]{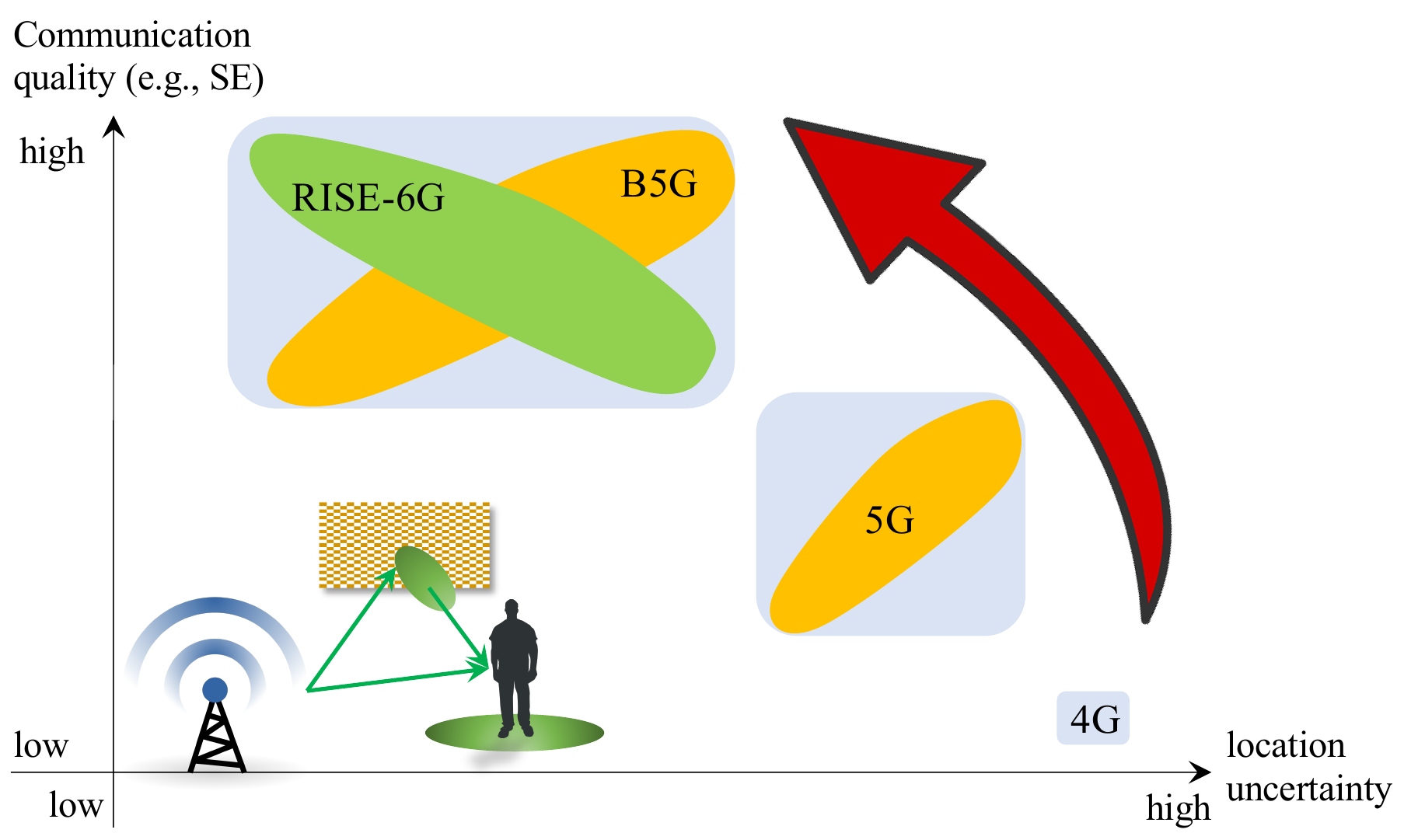}
    \caption{Localization and communication in RISE-6G will reinforce each other rather than compete for the same resources. 
    }
    \label{fig:WP45-figure}
\end{figure}
\subsection{RIS for Enhanced Localization and Sensing}
Localization and sensing 
are important enablers for the RISE-6G vision: not only do they have inherent value in a wide range of location-based services and applications, but they can also support enhanced connectivity, highly localized services' provision, 
as well as  opportunistic environmental and context awareness. 
As shown in Fig.~\ref{fig:WP45-figure}, it is our vision that in RISE-6G we see localization and communication as mutually reinforcing, rather than competing for the same resources. 
Several fundamental research questions still need to be addressed in this field of RIS-aided localization, mapping, and sensing, in addition to the issues already faced by RIS-enabled communications (e.g., RIS EM and channel modeling, 
channel estimation). First of all is the issue of localization-oriented RIS control: where should RIS be placed, how should they be controlled and optimized, at which time scales, in order to provide optimal localization accuracy and/or coverage? Secondly, given a chosen RIS-based system architecture, along with suitable RIS control methods, how do we estimate the users’ locations and possibly their orientations, in a resilient (e.g., against radio obstructions), highly flexible (e.g., on-demand) and scalable way (without necessitating re-deployment or re-planning), with 
minimum footprint in terms of latency, overhead, and overall consumption?
To address these questions, RISE-6G aims at developing localization-oriented architectures and control methods, operating jointly at RIS and system levels, benefiting from highly directional RIS operations, overall multipath channel reconfigurability, and a combination of model-based and data driven processing tools. Similarly to data communications, security, and low EMF, we envision the use of localization-boosted areas, where high performance could be guaranteed. We will also devise novel algorithms for positioning, tracking, as well active and passive detection, which relax the synchronization and overhead requirements inherent to 
conventional 
localization. 
This should lead to higher accuracy, service continuity, and flexibility than non-RIS solutions.

\subsection{RIS for Enhanced Sustainability and Security}
Enhancements in terms of sustainability and security will be attained by exploiting a novel and disruptive approach in the design and deployment of RIS-empowered networks, taking into account non-intended users such as EMF exposed users and eavesdroppers. This can be accomplished by leveraging sensing and EM-field 3D maps acquisitions enabled by RISs. We propose to feed the envisioned network with more knowledge on the non-intended users. Ideally, one would make a RIS smart enough to detect human bodies and configure itself to actively avoid EMF exposure (or eavesdropping) on them.

\begin{figure}
    \centering
    \includegraphics[width=1\columnwidth]{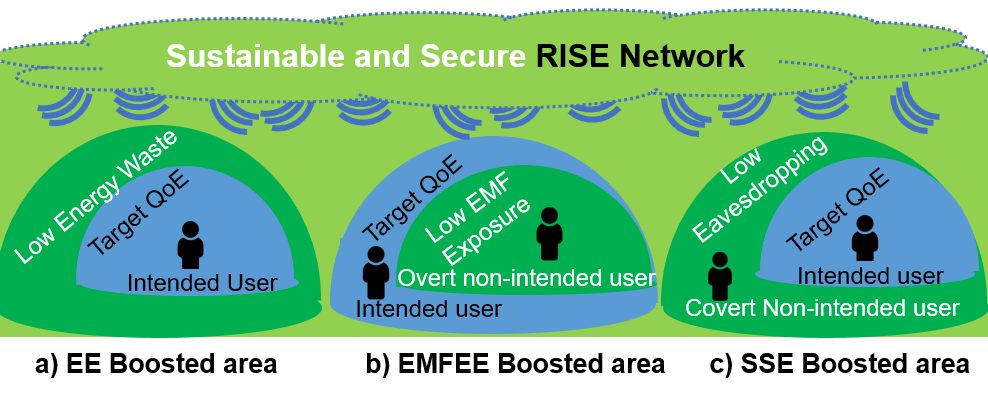}
    \caption{Illustrating the concepts of EE-boosted, EMFEE-boosted, and SSE-boosted areas, as well as the intended users and the overt and covert non-intended users.}
    \label{fig:mylabel}
\end{figure}

More precisely, we propose to jointly optimize RIS and beamforming in terms of EMFEE and SSE in addition to the more traditional EE metric. We will compare different levels of available knowledge for non-intended users. On one hand, EMF exposed users can fall in the category of \textit{overt} non-intended users, i.e. users that want to be identified as non-intended users so that the RIS-empowered network avoids them. Such overt non-intended users may cooperate to provide knowledge. On the other hand, eavesdroppers fall in the category of \textit{covert} non-intended users, i.e. users that do not want to be discovered; no cooperation is expected from such users. Thi terminology is illustrated in Fig.~\ref{fig:mylabel}. The level of signal at non-intended users will be taken into account as a constraint for the design of the active transmitting nodes' beamforming and the RIS tunable configurations. We will also devise coordinating beamforming algorithms and compare centralized and distributed solutions. We will consider solutions based on deep learning using cutting-edge neural network structures (e.g., generative adversarial networks). We will also propose new schemes to optimize the trade-off between the efficiency metrics, in the particular cases where EMFEE (or SSE) is improved at the expense of EE. 

Finally, RISE-6G will use the beyond state-of-the-art physics-based modeling and simulation tools for RISs, described in Sec.~\ref{sec:modeling}, to assess RIS-empowered networks in terms of EE, EMFEE, and SSE, taking into account the non-intended users in the assessment in various propagation environments. Measurements with a commercial 5G base station connected to the live network of Telecom Italia \cite{Dmicheli} will also pave the way for in-situ tests at the field trial facilities.

\section{RIS-Empowered Trials for 6G Use Cases}\label{sec:trial}
Realistic use cases will substantiate the attainable performance of RIS-empowered environments, while pursuing different objectives, such as overall performance optimization and localization accuracy maximization. This will be performed and hereafter described by means of two different field-trials that cover public and private scenarios: \textit{i}) a public train station; and \textit{ii}) a private smartly connected factory.

\subsection{Use Case 1: Train Station}
\begin{figure}
    \centering
    \includegraphics[width=1\columnwidth]{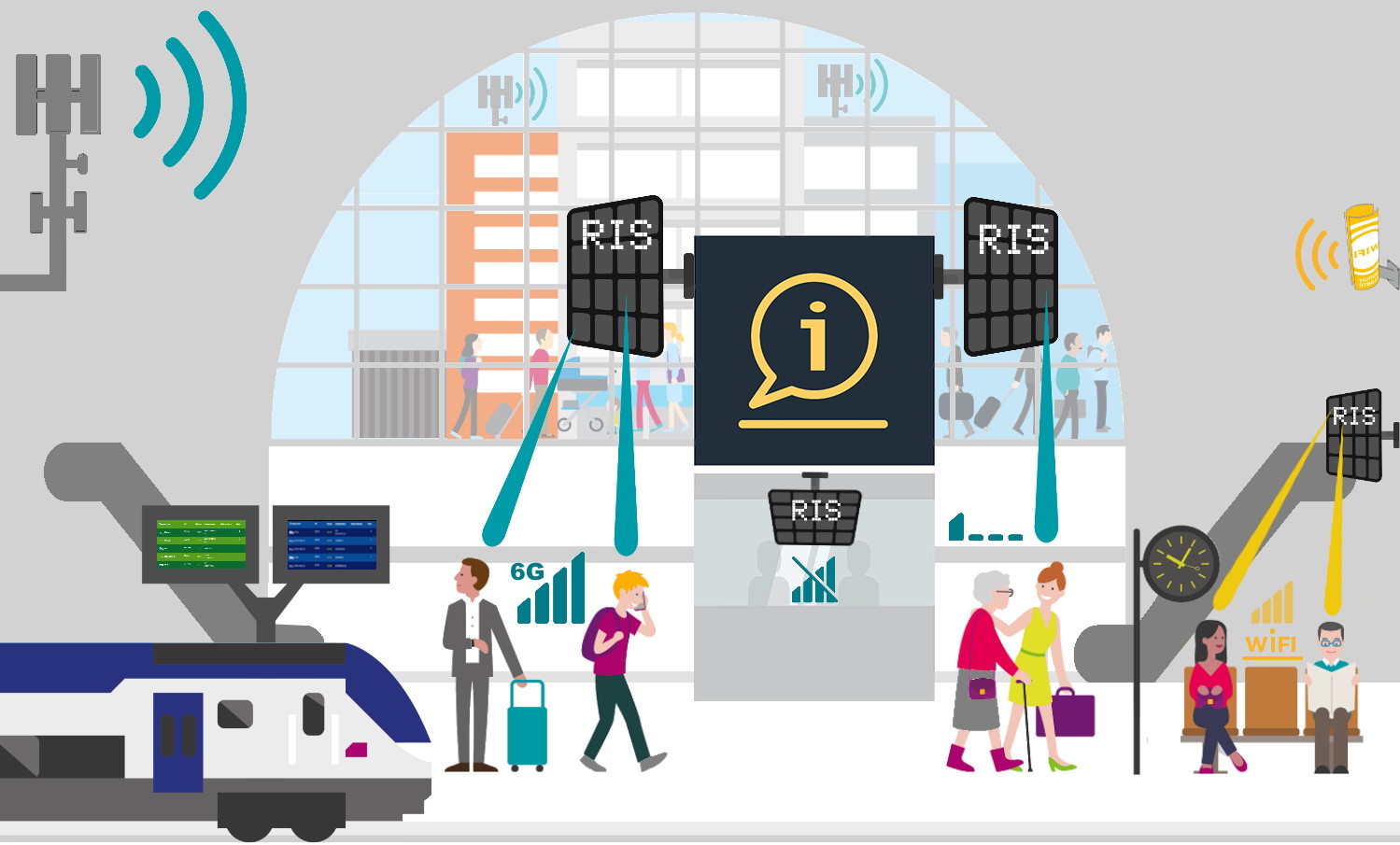}
    \caption{Envisioned RIS-empowered enhanced connectivity field trial at the Rennes, France train station.}\label{fig:train}
\end{figure}

An RIS-empowered solution will be properly and carefully set up in the train station located in Rennes, France, as depicted in Fig.~\ref{fig:train}. To facilitate the deployment of such proof-of-concept, generic distribution control networks will be developed to centrally control all RIS-based equipment considered in the scenario. We will identify key validation points as the following: \textit{i}) interfacing with RISs in terms of maximal density of independent channels over each RIS interface; and \textit{ii}) cost, power consumption, and complexity scaling functions with desired system performance.
 
Three relevant use cases have been identified that would greatly benefit from an RIS-empowered solution: \textit{i}) coverage optimization for indoor shops; \textit{ii}) dedicated resting areas for instant video download and cloud gaming ensuring the highest bandwidth and network efficiency; and \textit{iii}) EMF limitations in private areas or for workers protection purposes. For High Definition (HD) video connections, an RIS-empowered network will work at $26$ GHz by means of station infrastructure. At $70$ GHz, short-range high data rate video streaming will be tested with a compact RIS-based reflector, and a $77$ GHz radio module to support at least $1.488$ Gbps uncompressed HD video streaming with commercial video encoder 4K transmission will be demonstrated.

\subsection{Use Case 2: Connected Factory}
The RISE-6G project will deploy RISs within a smartly connected factory scenario. In particular, it will focus on the industrial operations in an automotive plant of Fiat-Chrysler Automobiles (FCA). Inside the plant, a variety of processes (logistics, for example, for sub-assembly, preparation, and transport of components or group of components (called as kits) from logistics areas to lineside) require precise positioning of their moving assets: components, containers, racks, kit handlers, Autonomous Guided Vehicles (AGV), unmanned aerial vehicles, and robots. In the case of the logistics process of kitting, the composition of an assembly kit includes: \textit{i}) containers of different components brought to a common area located near the final assembly line, stored and moving through the area; and \textit{ii}) components related to a single final product (e.g. vehicle, engine) picked and housed together, forming a kit. 

The state-of-the-art defines operations performed in a mixed way, partially automated (via AGVs) and partially manual (via human operators) so that the complete kit is brought to the assembly line in an automated way (via AGVs). The RISE-6G advanced solutions will deploy RISs to enable a complete mapping of the area (e.g., components, containers) in order to coordinate the AGVs' motion through continuous communication, facilitating the mission management. In particular, the component and container precise positioning and location in the rack, are not fixed on beforehand (that is, when the mission is given to the AGV) and might lead to a mission failure and a in-situ action of a logistics operator. In order to avoid potential failures it is, thus, crucial to achieve extreme accuracy in the localization process of the components, containers, or AGVs that must be carried out by means of advanced indoor localization techniques that cannot rely on cameras and distributed sensors. A novel framework to automatically control signal propagation properties by means of RISs will be delivered. Such RISs will complement the existing wireless network and enable a continuous communication channel between access points and the containers/AGVs via sub-6GHz and Ka-band ($26$ GHz) links. Additionally, cm-level localization demonstrations will be performed at $70$ GHz and $130$ GHz carrier frequencies. 
Finally, the proposed solution will target at minimizing the number of installed RISs required to seamlessly cover entire working areas, as well as at identifying the best localization algorithms offering resolution/refresh rate trade-offs.
\begin{figure}
    \centering
    \includegraphics[width=0.8\columnwidth]{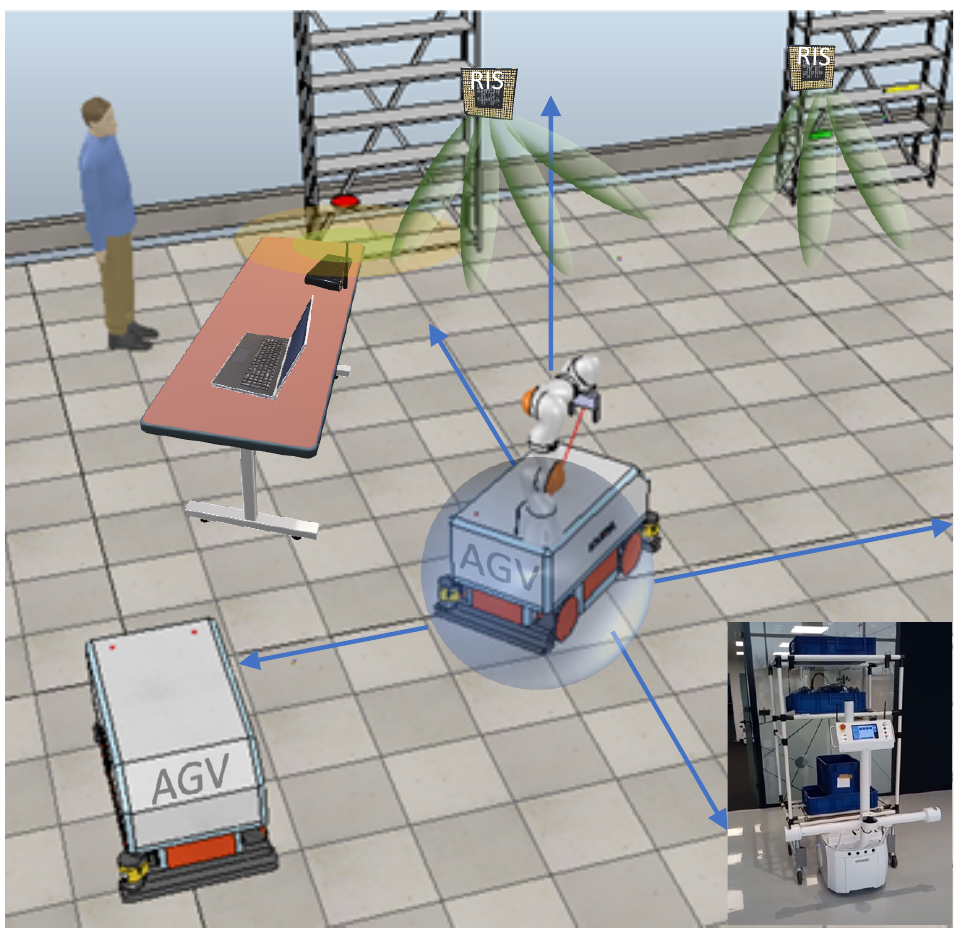}
    \caption{Envisioned RIS-empowered localization of AGVs equipped with a camera in kitting operations in a FCA automotive plant.}\label{fig:factory}
\end{figure}

\section{Conclusions}
This paper thoroughly overviewed the RISE-6G vision which capitalizes on the emerging RIS technology. The main technical challenges to properly tackle the unprecedented 6G network design requirements via the project's perspective were described: \textit{i}) definition of a novel network architecture and operation strategies incorporating multiple RISs; \textit{ii}) characterization of the fundamental limits of RIS-empowered communications, enabled by our proposed realistic and experimentally validated radio wave propagation models; \textit{iii}) design of efficient solutions to enable online trading between high-capacity connectivity, EE, EMFE, and localization accuracy based on dynamically programmable wireless propagation environments, while accommodating specific legislation and regulation requirements on spectrum use, data protection, and EMF emission; and \textit{iv}) prototype-benchmark proposed innovation via two complementary field trials with verticals.
 
\section{Acknowledgement}
This work has been supported by H2020 RISE-6G project.

\balance
\bibliographystyle{IEEEtran}
\bibliography{references}

\begin{thebibliography}{10}
\providecommand{\url}[1]{#1}
\csname url@samestyle\endcsname
\providecommand{\newblock}{\relax}
\providecommand{\bibinfo}[2]{#2}
\providecommand{\BIBentrySTDinterwordspacing}{\spaceskip=0pt\relax}
\providecommand{\BIBentryALTinterwordstretchfactor}{4}
\providecommand{\BIBentryALTinterwordspacing}{\spaceskip=\fontdimen2\font plus
\BIBentryALTinterwordstretchfactor\fontdimen3\font minus
  \fontdimen4\font\relax}
\providecommand{\BIBforeignlanguage}[2]{{%
\expandafter\ifx\csname l@#1\endcsname\relax
\typeout{** WARNING: IEEEtran.bst: No hyphenation pattern has been}%
\typeout{** loaded for the language `#1'. Using the pattern for}%
\typeout{** the default language instead.}%
\else
\language=\csname l@#1\endcsname
\fi
#2}}
\providecommand{\BIBdecl}{\relax}
\BIBdecl

\bibitem{Parkvall2020}
S.~Parkvall, Y.~Blankenship \emph{et~al.}, ``{5G NR} release 16: Start of the
  {5G} evolution,'' \emph{IEEE Commun. Standards Mag.}, vol.~4, no.~4, pp.
  56--63, Dec. 2020.

\bibitem{Calvanese2020}
E.~Calvanese~Strinati and S.~Barbarossa, ``{6G} networks: Beyond {S}hannon
  towards semantic and goal-oriented communications,'' \emph{J. Commun. Netw.},
  to appear, 2021.

\bibitem{Calvanese2019}
E.~Calvanese~Strinati, S.~Barbarossa \emph{et~al.}, ``{6G}: The next frontier:
  From holographic messaging to artificial intelligence using subterahertz and
  visible light communication,'' \emph{IEEE Veh. Technol. Mag.}, vol.~14,
  no.~3, pp. 42--50, Sep. 2019.

\bibitem{huang2019reconfigurable}
C.~Huang, A.~Zappone \emph{et~al.}, ``Reconfigurable intelligent surfaces for
  energy efficiency in wireless communication,'' \emph{IEEE Trans. Wireless
  Commun.}, vol.~18, no.~8, pp. 4157--4170, Aug. 2019.

\bibitem{liaskos2018new}
C.~Liaskos, S.~Nie \emph{et~al.}, ``A new wireless communication paradigm
  through software-controlled metasurfaces,'' \emph{IEEE Commun. Mag.},
  vol.~56, no.~9, pp. 162--169, Sep. 2018.

\bibitem{di2019smart}
M.~Di~Renzo \emph{et~al.}, ``Smart radio environments empowered by
  reconfigurable {AI} meta-surfaces: an idea whose time has come,''
  \emph{EURASIP J. Wireless Commun. Net.}, vol. 2019, no.~1, pp. 1--20, May
  2019.

\bibitem{huang2019holographic}
C.~Huang, S.~Hu \emph{et~al.}, ``Holographic {MIMO} surfaces for 6{G} wireless
  networks: Opportunities, challenges, and trends,'' \emph{IEEE Wireless
  Commun.}, vol.~27, no.~5, pp. 118--125, Oct. 2020.

\bibitem{WavePropTCCN}
G.~C. Alexandropoulos, G.~Lerosey \emph{et~al.}, ``Reconfigurable intelligent
  surfaces and metamaterials: {T}he potential of wave propagation control for
  {6G} wireless communications,'' \emph{IEEE ComSoc TCCN Newslett.}, vol.~6,
  no.~1, pp. 25--37, Jun. 2020.

\bibitem{DMA2020}
N.~Shlezinger, G.~C. Alexandropoulos \emph{et~al.}, ``Dynamic metasurface
  antennas for {6G} extreme massive {MIMO} communications,'' \emph{IEEE
  Wireless Commun.}, to appear, 2021.

\bibitem{hardware2020icassp}
G.~C. Alexandropoulos and E.~Vlachos, ``A hardware architecture for
  reconfigurable intelligent surfaces with minimal active elements for explicit
  channel estimation,'' in \emph{Proc. IEEE ICASSP}, Barcelona, Spain, May
  2020, pp. 9175--9179.

\bibitem{DiRenzoApr2020}
M.~{Di Renzo}, A.~{Zappone} \emph{et~al.}, ``Smart radio environments empowered
  by reconfigurable intelligent surfaces: How it works, state of research, and
  road ahead,'' \emph{IEEE J. Sel. Areas Commun.}, vol.~38, no.~11, pp.
  2450--2525, Nov. 2020.

\bibitem{risTUTORIAL2020}
Q.~Wu, S.~Zhang \emph{et~al.}, ``Intelligent reflecting surface aided wireless
  communications: {A} tutorial,'' \emph{IEEE Trans. Commun.}, to appear, 2021.

\bibitem{Gradoni2020}
G.~Gradoni and M.~Di~Renzo, ``End-to-end mutual-coupling-aware communication
  model for reconfigurable intelligent surfaces: An electromagnetic-compliant
  approach based on mutual impedances,'' \emph{IEEE Wireless Commun. Lett.}, to
  appear, 2021.

\bibitem{JSAC2020risma}
P.~{Mursia}, V.~{Sciancalepore} \emph{et~al.}, ``{RISMA}: Reconfigurable
  intelligent surfaces enabling beamforming for {IoT} massive access,''
  \emph{IEEE J. Sel. Areas Commun.}, to appear, 2021.

\bibitem{ZhangOFDM}
B.~Zheng and R.~Zhang, ``Intelligent reflecting surface-enhanced {OFDM}:
  Channel estimation and reflection optimization,'' \emph{IEEE Wireless Commun.
  Lett.}, vol.~9, no.~4, pp. 518--522, Apr. 2020.

\bibitem{HuangDNN}
C.~Huang, G.~C. Alexandropoulos \emph{et~al.}, ``Indoor signal focusing with
  deep learning designed reconfigurable intelligent surface,'' in \emph{Proc.
  IEEE SPAWC}, Cannes, France, Jul. 2019.

\bibitem{del2017survey}
J.~A. del Peral-Rosado, R.~Raulefs \emph{et~al.}, ``Survey of cellular mobile
  radio localization methods: From {1G} to {5G},'' \emph{IEEE Commun. Surveys
  Tuts.}, vol.~20, no.~2, pp. 1124--1148, 2017.

\bibitem{wymeersch2020radio}
H.~Wymeersch, J.~He \emph{et~al.}, ``Radio localization and mapping with
  reconfigurable intelligent surfaces: Challenges, opportunities, and research
  directions,'' \emph{IEEE Veh. Technol. Mag.}, vol.~15, no.~4, pp. 52--61,
  Dec. 2020.

\bibitem{Ma2020}
S.~Ma, S.~Phang \emph{et~al.}, ``Efficient statistical model for predicting
  electromagnetic wave distribution in coupled enclosures,'' \emph{Phys. Rev.
  Applied}, vol.~14, p. 014022, Jul. 2020.

\bibitem{Gradoni20181}
G.~{Gradoni}, L.~R. {Arnaut} \emph{et~al.}, ``Wigner-function-based propagation
  of stochastic field emissions from planar electromagnetic sources,''
  \emph{IEEE Trans. Electromagn. Compat.}, vol.~60, no.~3, pp. 580--588, Jun.
  2018.

\bibitem{Mudonhi21}
A.~Mudonhi, M.~Lotti \emph{et~al.}, ``{RIS}-enabled mmwave channel sounding
  based on electronically reconfigurable transmitarrays,'' in \emph{Proc. IEEE
  EuCAP}, Düsseldorf, Germany, Mar. 2021.

\bibitem{Dmicheli}
D.~Micheli, M.~Barazzetta \emph{et~al.}, ``Power boosting and compensation
  during {OTA} testing of a real {4G LTE} base station in reverberation
  chamber,'' \emph{IEEE Trans. Electromagn. Compat.}, vol.~57, no.~4, pp.
  623--634, Aug. 2015.

\end{thebibliography}
\end{document}